\documentstyle[12pt]{article}

\newcommand{\beq}{\begin{equation}}
\newcommand{\eeq}{\end{equation}}

\newcommand{\remove}[1]{}

\newcommand{\zbb}{$Zb\overline{b}$\ }
\newcommand{\bsg}{$b\rightarrow s \gamma$\ }

\begin{document}
\begin{titlepage}
\begin{center}
\today     \hfill    LBL-36217 \\
           \hfill    BUHEP-94-28\\

\vskip .5in

{\large \bf \mbox{\boldmath $b \rightarrow s \gamma$} and
\mbox{\boldmath $Z \rightarrow b \overline{b}$} in Technicolor with
Scalars }\footnote{This work was supported by the
Director, Office of Energy Research, Office of High Energy and Nuclear
Physics, Division of High Energy Physics of the U.S. Department of Energy
under Contract DE-AC03-76SF00098.}

\vskip 0.3in

Christopher D. Carone

{\em Theoretical Physics Group\\
    Lawrence Berkeley Laboratory\\
      University of California\\
    Berkeley, California 94720}

\vskip 0.3in

Elizabeth H. Simmons and Yumian Su

{\em    Boston University \\
        Department of Physics \\
        590 Commonwealth Avenue\\
        Boston, MA 02215}

\end{center}

\vskip .25in

\begin{abstract}
We consider the radiative decay $b\rightarrow s \gamma$, and the
correction to the $Zb\overline{b}$ vertex in technicolor models with
scalars.  In these models, the scalar develops a vacuum expectation value
when the technifermions condense, and the ordinary fermions develop masses
via Yukawa couplings.  Since the symmetry breaking
sector involves both a fundamental scalar doublet and an isotriplet of
composite scalars (the technipions), the phenomenology associated with
the charged scalars is similar to that found in a type-I two-Higgs
doublet model.  We show that the correction to the $Zb\overline{b}$ vertex
is small over the allowed parameter space of the model in the two limits
that we consider, and that there can be large, potentially observable,
contributions to the $b\rightarrow s \gamma$ branching fraction.
\end{abstract}
\end{titlepage}
\renewcommand{\thepage}{\roman{page}}
\setcounter{page}{2}
\mbox{ }

\vskip 1in

\begin{center}
{\bf Disclaimer}
\end{center}

\vskip .2in

\begin{scriptsize}
\begin{quotation}
This document was prepared as an account of work sponsored by the United
States Government. While this document is believed to contain correct
 information, neither the United States Government nor any agency
thereof, nor The Regents of the University of California, nor any of their
employees, makes any warranty, express or implied, or assumes any legal
liability or responsibility for the accuracy, completeness, or usefulness
of any information, apparatus, product, or process disclosed, or represents
that its use would not infringe privately owned rights.  Reference herein
to any specific commercial products process, or service by its trade name,
trademark, manufacturer, or otherwise, does not necessarily constitute or
imply its endorsement, recommendation, or favoring by the United States
Government or any agency thereof, or The Regents of the University of
California.  The views and opinions of authors expressed herein do not
necessarily state or reflect those of the United States Government or any
agency thereof or The Regents of the University of California and shall
not be used for advertising or product endorsement purposes.
\end{quotation}
\end{scriptsize}

\vskip 2in

\begin{center}
\begin{small}
{\it Lawrence Berkeley Laboratory is an equal opportunity employer.}
\end{small}
\end{center}

\newpage
\renewcommand{\thepage}{\arabic{page}}
\setcounter{page}{1}

\section {Introduction} \label {sec:intro}

The phenomenology of technicolor models with scalars has been considered
extensively in the recent literature \cite{sim,ksam,cg,cs,cgold}.  In this
class of models, the technifermions and the ordinary fermions both couple to
a weak scalar doublet, which replaces the conventional ETC sector.  When
technicolor becomes strong, and a technifermion condensate forms, the Yukawa
coupling of the condensate to the scalar produces a linear term in the scalar
potential.  As a result, the scalar develops a vacuum expectation
value (vev), which is responsible for giving the ordinary fermions mass.
It has been shown that models of this type do not produce unacceptably
large contributions to $K^0$-$\overline{K^0}$ or $B^0$-$\overline{B^0}$
mixing, nor to the electroweak $S$ and $T$ parameters \cite{sim,cg,cs}.
In addition, the new scalars in the model can be made heavy enough to evade
detection, even in the limit where the scalar doublet is assumed to have a
vanishing $SU(2) \times U(1)$ invariant mass \cite{cg}. Technicolor with
scalars is interesting on more general grounds because it can arise as a
low-energy effective theory in strongly-coupled ETC (SETC) models
\cite{setc}, providing that some degree of fine-tuning is allowed.  This
fine-tuning is necessary in any workable SETC model to maintain a sufficient
hierarchy between the ETC and technicolor scales \cite{kl}.

It is the purpose of this letter to consider two other important
phenomenological issues that have not been studied in the context
of technicolor models with scalars: the correction to the \zbb coupling,
and the branching fraction B($b\rightarrow s \gamma$).  In conventional
ETC models, it has been shown that there is a reduction in the \zbb coupling
proportional to $m_t$, which can decrease
$R_b \equiv \Gamma(Z\rightarrow b\overline{b}) /
\Gamma(Z\rightarrow {\rm hadrons})$ by as much as 5\% \cite{scetal}.  In
two-Higgs doublet models, and in the minimal supersymmetric standard model,
the \zbb coupling can shift as a consequence of radiative corrections
involving charged scalars that couple $b$ to $t$ through the large top-quark
Yukawa coupling.  Technicolor with scalars is similar to a type-I
two-Higgs doublet model, in which only one scalar doublet (in our case, the
fundamental scalar) couples to both the charge 2/3 and charge -1/3 quarks.
Unlike the situation in conventional ETC models, the main correction
to the \zbb coupling in technicolor models with scalars comes from
the radiative effects of the physical charged scalars in the
low-energy theory.  It is therefore possible to adapt much of the
existing analysis of the \zbb coupling in two-Higgs doublet models
to study the parameter space in the models of interest to us here.

The other process that we consider, \bsg, vanishes at tree-level in the
Standard Model due to the GIM mechanism, but can occur at one-loop through
penguin diagrams.  Since the Standard Model branching fraction depends on
small GIM-violating effects, it has been suggested that this decay mode may
provide a sensitive probe of physics beyond the Standard Model \cite{probe}.
The recent measurement by the CLEO collaboration of the inclusive
\bsg decay width has yielded the bound $M_{H^+} > 260$ GeV for the charged
scalar mass in type-II two-Higgs doublet models \cite{cleo}.  This gives us
substantial motivation to study whether \bsg can receive important
contributions in technicolor models with scalars.  Again, existing
calculations of the inclusive decay width in type-I two-Higgs doublet
models can be modified to enable us to determine the \bsg
width throughout our model's parameter space.

\section{The Model}

The model that we consider has been described in detail
elsewhere \cite{sim,cg}, so we summarize only the essential components.
In addition to the Standard Model gauge structure and particle content, we
assume  a minimal $SU(N)$ technicolor sector, with two techniflavors that
transform as a left-handed doublet and two right-handed singlets under
$SU(2)_W$,
\beq
\Upsilon_L=\left(\begin{array}{c} p \\ m
\end{array} \right)_L  \,\,\,\,\,p_R \,\,\,\, m_R
\eeq
and that have the weak hypercharge assignments $Y(\Upsilon _L)=0$,
$Y(p_R)=1/2$, and $Y(m_R)=-1/2$. The ordinary fermions are technicolor
singlets, with their usual quantum number assignments.  The technifermions
and ordinary fermions couple to a weak scalar doublet which has the quantum
numbers of the Higgs doublet of the Standard Model
\beq
\phi=\left(\begin{array}{c} \phi^+ \\ \phi^0
\end{array}\right)
\eeq
The purpose of the scalar is to couple the technifermion condensate
to the ordinary fermions and thereby generate fermion masses.

What is relevant to the phenomenology that we consider here is that the
technipions (the isotriplet scalar bound states of $p$ and $m$) and the
isotriplet components of $\phi$ will mix.  One linear combination becomes
the longitudinal component of the $W$ and $Z$.  The orthogonal linear
combination remains in the low-energy theory as an isotriplet of physical
scalars.  Denoting the physical scalars $\pi_p$, the coupling of the
charged physical scalars to the quarks is given by \cite{cg}
\beq
i(\frac{f}{v})\left[ \overline{D_L} V^\dagger \pi^-_p h_U U_R
+\overline{U_L} \pi^+_p V h_D D_R + h.c.\right]
\label{cpcoup}
\eeq
where $V$ is the Cabibbo-Kobayashi-Maskawa (CKM) matrix, $f$ is the
technipion decay constant, and $v$ is the electroweak scale
$\approx 250$ GeV.  Here $U$ and $D$ are column vectors in flavor
space, and the Yukawa coupling matrices are diagonal $h_U=diag(h_u,h_c,h_t)$,
$h_D=diag(h_d,h_s,h_b)$.  Notice that (\ref{cpcoup}) has the same form as
the charged scalar coupling in a type-I two-Higgs doublet model.

The fact that the quarks couple only to the fundamental scalar, but
not to the technipions, also accounts for the dependence of (\ref{cpcoup})
on $f/v$.   The technicolor scale and the scalar vev,
which we will call $f'$, both contribute to the electroweak scale
\beq
f^2+f'^2=v^2
\label{ews}
\eeq
In the limit that $f\rightarrow 0$, the fundamental scalar vev determines
the electroweak scale, and the longitudinal components of the weak gauge
bosons are mostly the fundamental scalar.  Thus, the physical charged
scalars are now mostly technipion in this limit, and we expect
their couplings to the quarks to vanish.  The couplings in (\ref{cpcoup})
have the correct behavior in this limit.

The chiral Lagrangian analysis in Refs.~\cite{cg,cs} allows us to estimate
the mass of the charged scalars.  At lowest order, the mass
of the physical isotriplet is given by
\beq
m_\pi^2=2c_1\sqrt{2}\frac{4\pi f}{f'} v^2 h
\label{mpip}
\eeq
where $h$ is the average technifermion Yukawa coupling
$h\equiv (h_++h_-)/2$, and where $h_+$ and $h_-$ are the individual Yukawa
couplings to $p$ and $m$, respectively.  The constant $c_1$ is an
undetermined coefficient in the chiral expansion, but is of order unity by
naive dimensional analysis (NDA) \cite{nda}.  We set $c_1=1$ in all
numerical estimates to follow.  Since we are only working to lowest order,
$c_1$ and $h$ always appear in the combination $c_1 h$; thus the uncertainty
in our estimate of $c_1$ can be expressed alternatively as an uncertainty in
the value of $h$.

The only remaining ingredient that we need for the
numerical analysis is the dependence of $f$ and $f'$ on the free parameters
of the model.  In general, $f$ and $f'$ can depend on $h_+$, $h_-$, $M_\phi$,
and $\lambda$, where $M_\phi$ is the $SU(2)\times U(1)$ invariant
scalar doublet mass, and $\lambda$ is the $\phi ^4$ coupling.  Two
limits have been studied previously in the literature:
{\em (i)} the limit in which $\lambda$ is small and can be neglected
\cite{sim}, and {\em (ii)} the limit in which $M_\phi$ is small and can be
neglected \cite{cg}.   One advantage of working in these two limits
is that the phenomenology of the model can be described simply in terms of a
two-dimensional parameter space, either ($M_\phi$,$h$) or ($\lambda$,$h$).
(This is possible because $h_+$ and $h_-$ enter only through the
combination $h=(h_++h_-)/2$ at lowest order in the chiral expansion.) In
general, the condition that the Higgs field $\sigma$ (the isoscalar component
of $\phi$) has no vacuum expectation value
\beq
V'(\sigma=0)=0
\eeq
gives us the constraint
\beq
M^2_\phi f' + \frac{\lambda}{2} {f'}^3 = 8 \sqrt{2} c_1 \pi
h f^3
\label{cons}
\eeq
where the last term is induced by the technicolor interactions \cite{cg,cs}.
In either limit {\em (i)} or {\em (ii)} described above, $f$ and $f'$ can
be found by solving (\ref{ews}) and (\ref{cons}) simultaneously in
terms of the two remaining free parameters.

If we also include the largest Coleman-Weinberg corrections to the
potential
\beq
V_{CW} = -\frac{1}{64\pi^2}\left(3h^4_t+2Nh^4\right) \sigma^4 \log
\left(\frac{\sigma^2}{\mu^2}\right)
\label{cw}
\eeq
(setting $h_+=h_-$ for simplicity) and define the renormalized $\phi^4$
coupling $\lambda_r=V''''(f')/3$ in order to remove the $\mu$-dependence
in (\ref{cw}), then the form of (\ref{cons}) will remain unchanged providing
that we work with the shifted parameters
\beq
\tilde{M}_\phi^2 = M_\phi^2 + \left(\frac{44}{3}\right)
\frac{1}{64\pi^2}\left[3h^4_t+2Nh^4\right] f'^2
\eeq
in limit {\em (i)}, or \cite{cg}
\beq
\tilde{\lambda}=\lambda + \frac{11}{24\pi^2}
\left[3h^4_t+2Nh^4\right]
\eeq
in limit {\em (ii)}. It is convenient for us to adopt these
new parameters because they absorb the effects of
the radiative corrections, as far as they affect the phenomenology
of the charged scalars. However, the isoscalar mass
$m_\sigma$ will have a different functional dependence on the
new parameters.  In limit {\em (i)}
\beq
m_\sigma^2 = \tilde{M}_\phi^2 + (\frac{64}{3})
\frac{1}{64\pi^2}\left[3h^4_t+2Nh^4\right] f'^2
\label{msigone}
\eeq
while in limit {\em (ii)}
\beq
m_\sigma^2 = \frac{3}{2}\tilde{\lambda} f'^2-
\frac{1}{8\pi^2}\left[3h^4_t+2Nh^4\right] f'^2
\label{msigtwo}
\eeq
In this paper we will again study limits {\em (i)} and {\em (ii)}, in
the interest of completing the phenomenological discussion presented in
Refs.~\cite{cg,cs}.

\section{Results}

Given the couplings in (\ref{cpcoup}), and estimating the charged scalar
mass, from equations (\ref{ews}), (\ref{mpip}) and (\ref{cons}), we can
apply previously published two-Higgs doublet model results to study our
model in limits {\em (i)} and {\em (ii)}.  The one-loop effects of charged
scalars on the \zbb coupling are discussed in Refs.~\cite{bfin,hol}.  We
adapt the results given in the appendix of Boulware and Finnel \cite{bfin}.
We compute the quantity $\delta R_b / R_b$, where $R_b$ is defined by
\beq
R_b = \frac{\Gamma (Z \rightarrow b \overline{b})}
{\Gamma(Z \rightarrow {\rm hadrons})}
\eeq
The contours of constant $\delta R_b / R_b$ are plotted in Figures
\ref{zbbh} and \ref{zbbl}, for the model in limits {\em (i)} and {\em (ii)}
respectively.  In both cases, we show the parameter space already
excluded by $B^0$-$\overline{B^0}$ mixing (the area below the
B-line), and by the constraint $m_{\pi^+}>m_t-m_b$ (the area to
the left and below the $m_\pi=m_t-m_b$ line).  We assume
a Standard Model top quark, with $m_t=175$ GeV and with no
decay to $\pi^+ b$, consistent with the recent CDF results \cite{cdfr}.
The $h f'= 4 \pi f$ line shown in Figure \ref{zbbh} indicates where the
chiral Lagrangian analysis breaks down; above this line, the technifermion
current masses are no longer small compared to the chiral symmetry breaking
scale, and we can make no claims about the phenomenology.  Note, however,
that this problematic region is avoided in limit {\em (ii)}, because this
area is already excluded by the constraint of vacuum stability,
$m_\sigma^2>0$ \cite{cg}.  The differing form of (\ref{msigone})
and (\ref{msigtwo}) explains why we obtain a strong constraint from the
the LEP lower bound \cite{lep} on the mass of the light neutral isosinglet
scalar in limit {\em (ii)}, but not in limit {\em (i)}. In limit {\em (i)},
one must go to negative values of $\tilde{M}_\phi$ before $m_\sigma$ becomes
in conflict with the LEP limit $m_\sigma>58.4$ GeV; however, the isotriplet
scalars become unacceptably light long before then.  In both limits we see
that the -1\% contour for $\delta R_b / R_b$ is almost contiguous with
the B-line, and that $\delta R_b / R_b$ becomes smaller as one moves away
from the excluded region in the lower right-hand portion of the plots.  Thus,
we see that a more dramatic effect corresponding to a larger top quark
Yukawa coupling is already precluded by the $B^0$-$\overline{B^0}$ mixing
constraints.

The partial width for \bsg in two-Higgs doublet models has
been computed in Refs.~\cite{gsw,mbsg} .  We adopt the results of
Grinstein, Springer, and Wise (GSW) \cite{gsw} in our analysis.
In Figures \ref{bsgh} and \ref{bsgl} we plot contours of constant
$\delta \Gamma / \Gamma$, the percent shift in the \bsg partial width
relative to the theoretical prediction in the Standard Model
(The Standard Model prediction corresponds to a branching fraction
of $2.75\times 10^{-4}$ \cite{cleo}).  In terms of the function $C_7$
defined in GSW, we plot contours of constant
\beq
\frac{(|C_7(m_b)_{2HD}|^2-|C_7(m_b)_{SM}|^2)}{|C_7(m_b)_{SM}|^2}
\label{c7ratio}
\eeq
where $2HD$ here refers to a type-I two-Higgs doublet model.
Notice that the contours are roughly parallel to the B-line,
and may eventually provide a tighter constraint. The 95\% confidence
level lower bound found by CLEO corresponds to the -64\% contour on
the plot, but this statement does not take into account the
theoretical uncertainty in (\ref{c7ratio}).  If we assume that
$C_7(m_b)$ can be calculated to within 15\%, as is suggested in \cite{gsw},
and we assume, conservatively, that (\ref{c7ratio}) is known within
30\%, then the region absolutely excluded in our model lies below the -83\%
contour.  This boundary is almost contiguous with the B-line, and
does not eliminate any additional parameter space. What is interesting
is that there is plenty of parameter space in which the
correction to \bsg is significant, (between -1\% and -50\%) yet not in
conflict with the lower bound on $\Gamma (b\rightarrow s \gamma)$ found by
CLEO.  In addition, the region of the parameter space in which the correction
to \bsg is less than -1\% includes the region discussed in
Ref.~\cite{cgold} which should include a light, extremely narrow technirho.

\section{Conclusions}
We have extended the phenomenological analysis of Refs.~\cite{sim,cg,cs,cgold}
to include corrections to \bsg and to the \zbb vertex in technicolor models
with scalars.   We have shown that the correction to $R_b$ is negative
throughout the allowed region of our model's parameter space in the limits
that we considered, but never larger than -1\%.  This result is consistent
with our expectation that corrections to the \zbb coupling are suppressed
in SETC models by an increased ETC scale \cite{evans}. Since the current
experimental measurement is {\em larger} than the Standard Model
prediction by two standard deviations \cite{pdg}, it may still be possible
to rule out this model if the Standard Model is ruled out on similar
grounds.  In addition, we have found sizable corrections to the \bsg width,
yielding a branching fraction that is smaller than the Standard Model
expectation.  Nevertheless, in both cases our results indicate that
stongly-coupled ETC models, of which our model is the low-energy limit,
easily survive the current experimental constraints.  Unlike the
electroweak $S$ and $T$ parameters, the \bsg branching fraction can be
calculated more reliably in these models, giving us relatively definitive
predictions.  Since the deviation from the Standard Model branching fraction
can be sizable in some regions of our model's parameter space, this effect
may be observable given improved measurements of the \bsg inclusive decay
width.

\begin{center}
{\bf Acknowledgments}
\end{center}
We thank Hitoshi Murayama and John March-Russell for helpful comments.
Simmons acknowledges the financial support of an American Fellowship from
the American Association of University Women. {\em This work was supported
by the Director, Office of Energy Research, Office of High Energy and
Nuclear Physics, Division of High Energy Physics of the U.S. Department of
Energy under Contract DE-AC03-76SF00098.}



\newpage
\begin{center}
{\bf Figure Captions}
\end{center}

\begin{figure}[h]
\caption{ \label{zbbh} Contours of constant $\delta R_b /R_b$
(dotted lines) in limit {\em (i).} The allowed region lies
above the B-line, and above the $m_\pi=m_t-m_b$ line.}
\caption{\label{zbbl} Contours of constant $\delta R_b /R_b$
(dotted lines) in limit {\em (ii).} The allowed region lies
above the B-line, above the $m_\pi=m_t-m_b$ line, and below the
$m_\sigma=58.4$ Gev line.}
\caption{\label{bsgh} Percent shift in the \bsg decay width
(dashed lines) in limit {\em (i)}, relative to the Standard Model
prediction. The -100\% contour is outside the allowed region,
but is provided for reference.}
\caption{\label{bsgl} Percent shift in the \bsg decay width
in limit {\em (ii)}, relative to the Standard Model prediction.}
\end{figure}


\begin{thebibliography}{99}
\frenchspacing

\bibitem{sim}
E.H. Simmons, Nucl. Phys. {\bf B312} 253 (1989).
\bibitem{ksam}
S. Samuel, Nucl. Phys. {\bf B347} 625 (1990); A. Kagan and S. Samuel,
Phys. Lett. {\bf B252} 605 (1990) and Phys. Lett. {\bf B270} 37 (1991).
\bibitem{cg}
C.D. Carone and H. Georgi, Phys. Rev. {\bf D49} 1427 (1994).
\bibitem{cs}
C.D. Carone and E.H. Simmons, Nucl. Phys. {\bf B397} 591 (1993).
\bibitem{cgold}
C.D. Carone and M. Golden, Phys. Rev. {\bf D49} 6211 (1994).
\bibitem{setc}
R.S. Chivukula, A.G. Cohen and K. Lane, Nucl. Phys. {\bf B343}
554 (1990); T. Appelquist, J. Terning, and L.C.R. Wijewardhana,
Phys. Rev. {\bf D44} 871 (1991).
\bibitem{kl}
K. Lane, BUHEP-94-24 (1994) (bboard:hep-ph@xxx.lanl.gov - 9409304).
\bibitem{scetal}
R.S. Chivukula, E. Gates, E.H. Simmons and J. Terning, Phys.
Lett. {\bf B311} 157 (1993); R.S. Chivukula, E.H. Simmons, J.
Terning, Phys. Lett. {\bf B331} 383 (1994).
\bibitem{probe}
J.L. Hewett, SLAC-PUB-6521 (1994).
\bibitem{cleo}
B. Barish, {\em et. al.}, CLEO Collaboration, CLEO CONF 94-1.
\bibitem{nda}
A. Manohar and H. Georgi, Nucl. Phys. {\bf B234} 189 (1984);
H. Georgi and L. Randall, Nucl. Phys. {\bf B276} 241 (1986);
H. Georgi, Phys. Lett. {\bf B298} 187 (1993).
\bibitem{bfin}
M. Boulware and D. Finnell, Phys. Rev. {\bf D44} 2054 (1991).
\bibitem{hol}
W. Hollik, Mod. Phys. Lett. {\bf A5} 1909 (1990).
\bibitem{cdfr}
CDF Collaboration, Phys. Rev. {\bf D50} 2966 (1994).
\bibitem{lep}
Buskulic, {\em et. al.}, Phys. Lett. {\bf B313} 299 (1993).
\bibitem{gsw}
B. Grinstein, R. Springer, and M.B. Wise, Nucl. Phys. {\bf B339}
269 (1990).
\bibitem{mbsg}
T.G. Rizzo, Phys. Rev. {\bf D38} 820 (1988);W.S. Hou and R.S. Willey,
Phys. Lett. {\bf B202} 591 (1988); C.Q. Geng and J.N. Ng, Phys.
Rev. {\bf D38} 2858 (1988); V. Barger, J.L. Hewett, and R.J.N.
Phillips, Phys. Rev. {\bf D41} 3421 (1990).
\bibitem{evans}
N. Evans, Phys. Lett. {\bf B331} 378 (1994).
\bibitem{pdg}
Particle Data Group, Phys. Rev. {\bf D50} 1172 (1994).
\end{thebibliography}
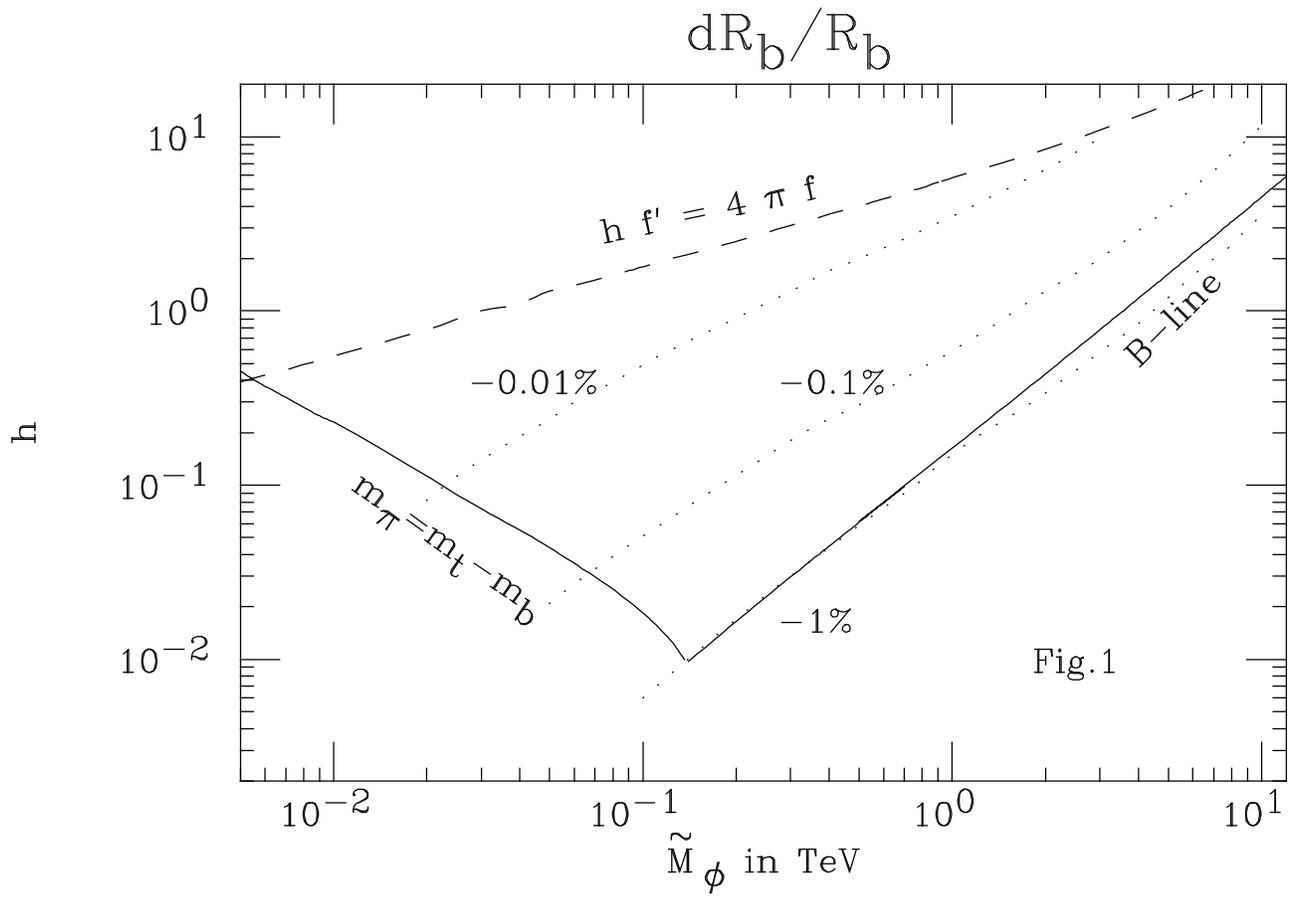
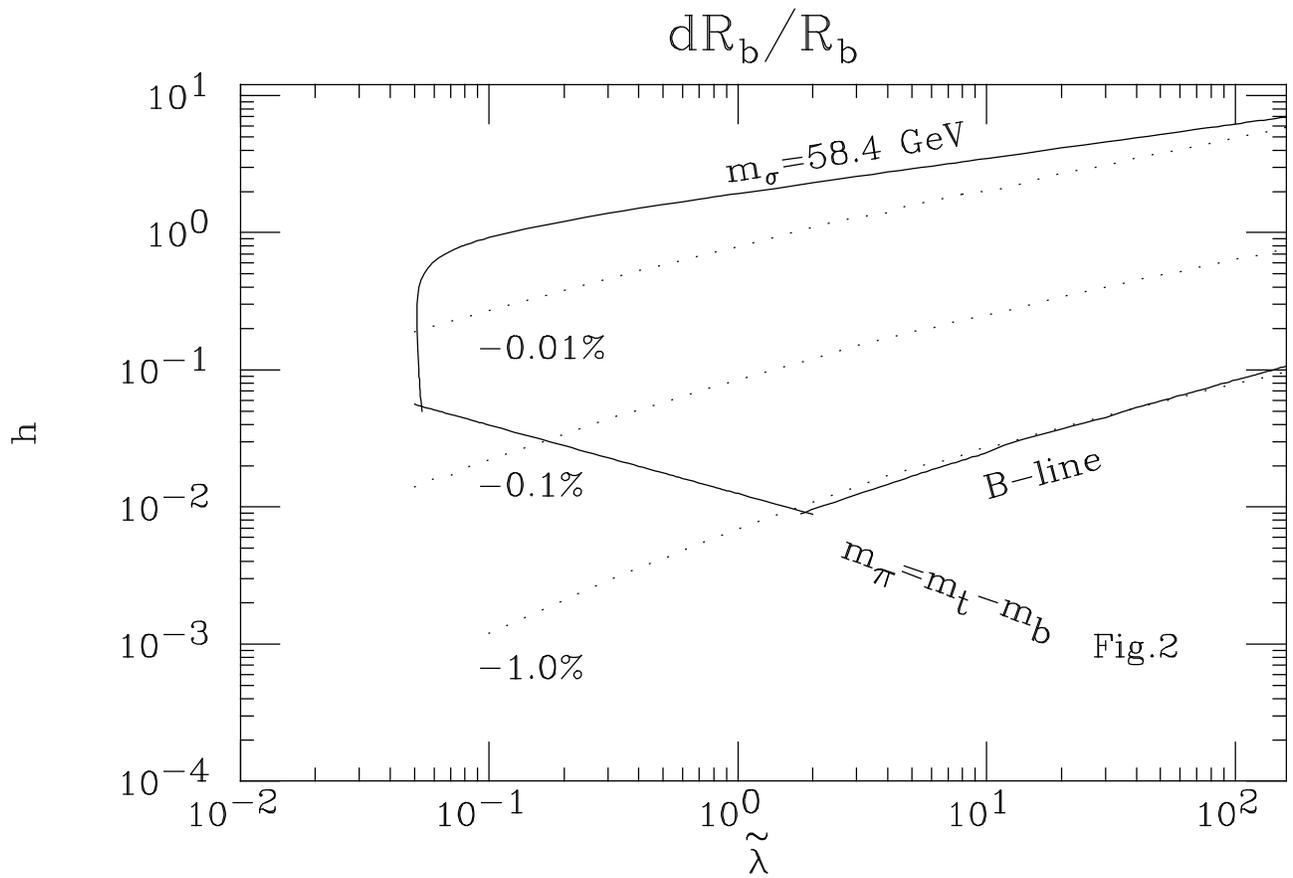
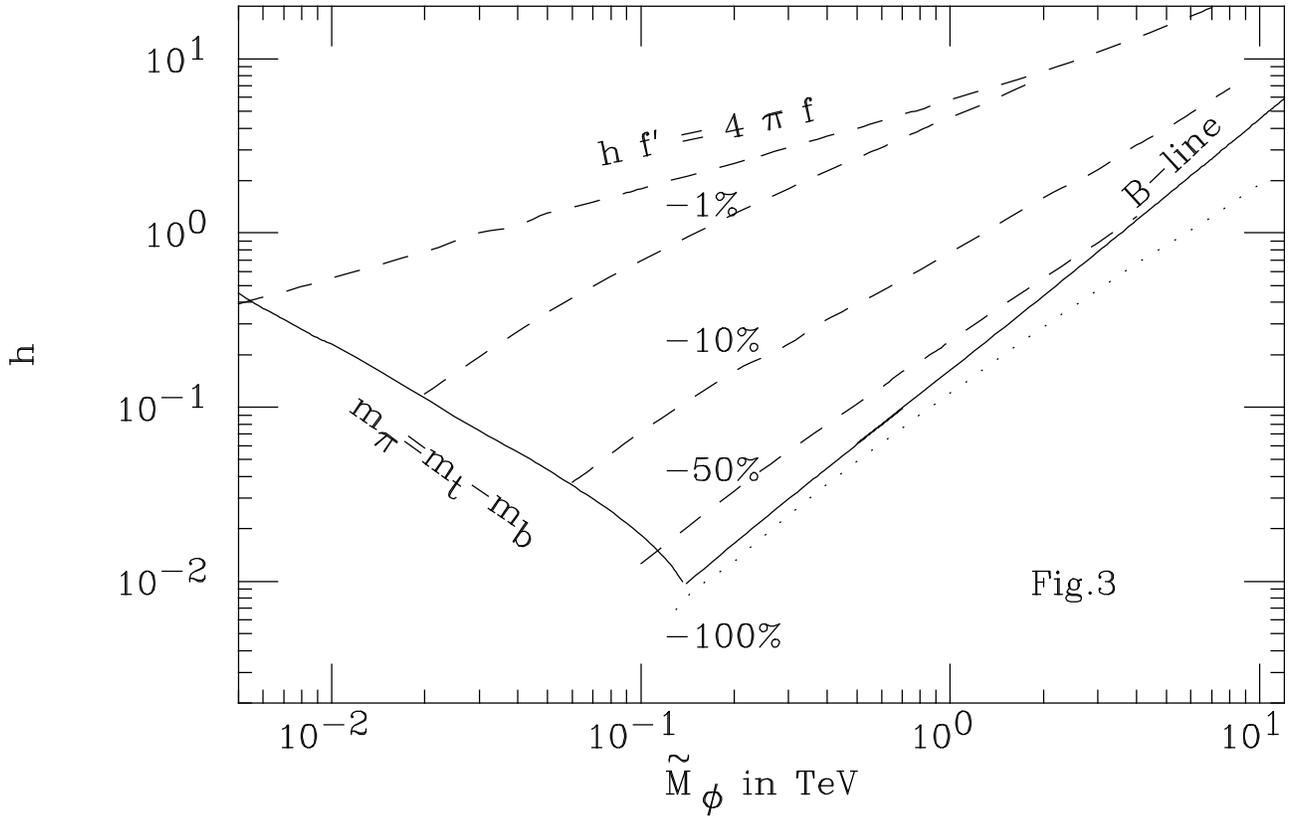
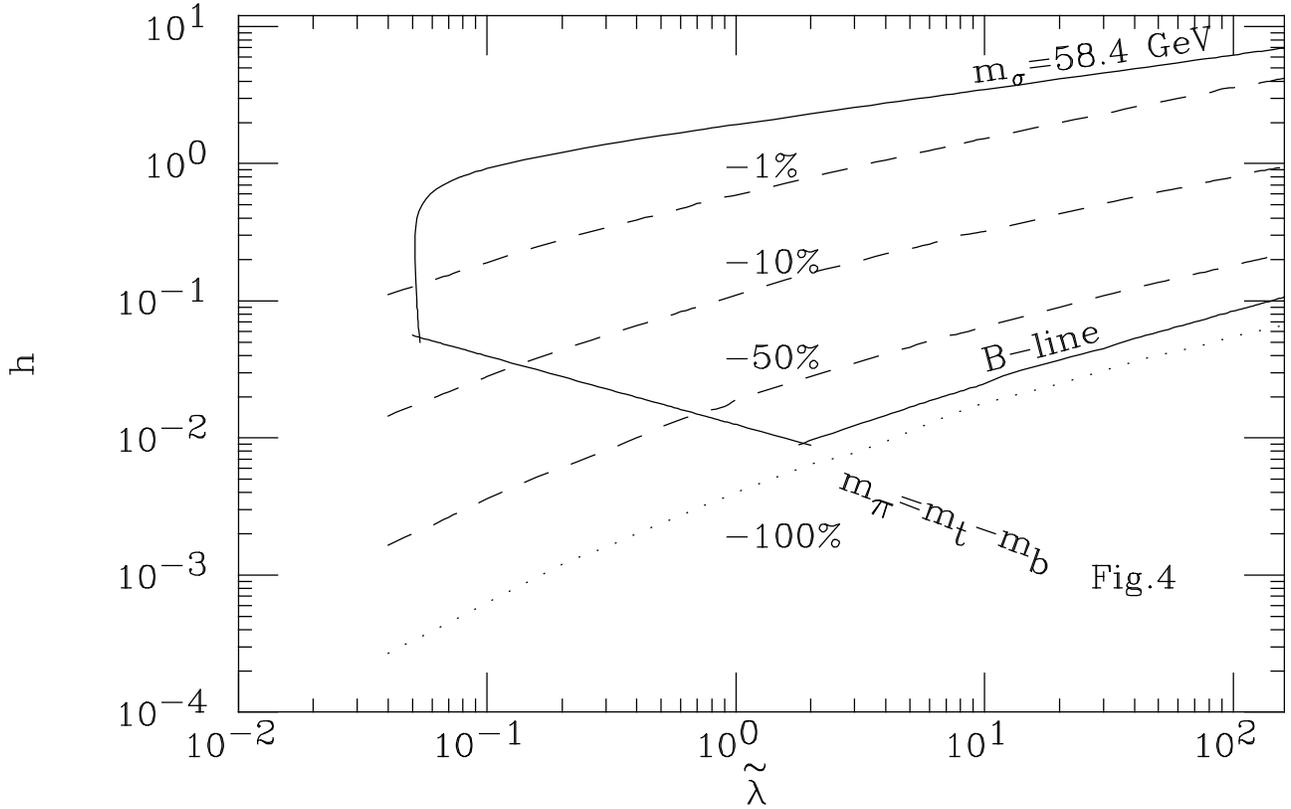
\end{document}